\documentclass[epj]{webofc}
\usepackage[varg]{txfonts}   
\usepackage{lineno}
\def\rp{{R}_{p}}

\def\lsim{\raise0.3ex\hbox{$\;<$\kern-0.75em\raise-1.1ex\hbox{$\sim\;$}}}
\def\gsim{\raise0.3ex\hbox{$\;>$\kern-0.75em\raise-1.1ex\hbox{$\sim\;$}}}

\def    \beq            {\begin{equation}}
\def    \eeq            {\end{equation}}
\def    \bea           {\begin{eqnarray}}
\def    \eea           {\end{eqnarray}}

\def \mn{\mu\nu{\rm SSM}}

\def\g2{{\rm GeV}^2}

\def\sw2{sin^2 \theta_w}

\def\a^tau{\alpha_{\tau}}

\def\beq{\begin{equation}}
\def\eeq{\end{equation}}
\def\beqa{\begin{eqnarray}}
\def\eeqa{\end{eqnarray}}

\newcommand{\newc}{\newcommand}
\newc\BR{BR}
\newc{\akappa}{A_{\kappa} }
\newc\deltagmtwo{\delta (g-2)_{\mu}} 
\newc\deltaamu{\Delta a_{\mu}}

\def\anti{\overline}

\def\rpv{{R}_{p} \hspace{-0.33cm}\slash\hspace{0.2cm}}

\newc{\haa}{BR\(h_1\to a_1 a_1\)}
\newc{\abb}{BR\(a_1\to b\anti{b}\)}
\newc{\hbb}{BR\(h_1\to b\anti{b}\)}
\newc{\Fermi}{\textit{Fermi}-}
\newc{\abund}{\Omega h^2}
\newc\bsgamma{b\rightarrow s \gamma }
\newc\bxsgamma{\overline{B}\rightarrow X_{s}\gamma}
\newc\brbsgamma{\BR(\overline{B}\rightarrow X_s\gamma)}

\begin{document}
\thispagestyle{empty}
\begin{flushright}
FTUAM 17/2\\
IFT-UAM/CSIC-17-005\\
\vspace*{5mm}
\end{flushright}

\vspace{-1.5cm}

%
%
\title{Models of Supersymmetry for Dark Matter}
%
%

\author{Carlos Mu\~noz\inst{1,2}\fnsep
\thanks{\email{c.munoz@uam.es}} 
}

\institute{Departamento de F\'{\i}sica Te\'{o}rica, Universidad Aut\'{o}noma de Madrid,
Cantoblanco, E-28049 Madrid, Spain
\and
Instituto de F\'{\i}sica Te\'{o}rica UAM-CSIC, Campus de Cantoblanco, E-28049 Madrid, Spain
          }

\abstract{%
A brief review of supersymmetric models and their candidates for dark matter is carried out. 
The neutralino is a WIMP candidate 
in the MSSM where $R$-parity is conserved, but this model has the $\mu$ problem. 
There are 
natural 
solutions to this problem that
necessarily introduce new structure beyond the MSSM, including new candidates for dark matter. In particular, in an extension of the NMSSM, the right-handed sneutrino can be used 
for this job.
In $R$-parity violating models such as the $\mn$, the gravitino
can be the dark matter, and could be detected by its decay products in gamma-ray experiments.
}
\maketitle
\section{Introduction}
\label{intro}

The Higgs particle in the standard model is intriguing, being the only elementary scalar in the spectrum, and introducing the hierarchy problem in the theory.
In supersymmetry (SUSY), the presence of the Higgs is more natural:  
scalar particles exist by construction, the hierarchy problem can be solved, and the models predict that the Higgs mass must be 
$\lsim$ 140 GeV if perturbativity of the relevant couplings up to high-energy scales is imposed. In a sense, the latter has been confirmed by the detection of a scalar particle with a mass of about 125 GeV.
However, in SUSY at least two Higgs doublets are necessary, and as a consequence new neutral and charged scalars should be detected in the future to confirm the theory. Not only that, as is well known, the spectrum of elementary particles is in fact doubled with masses of about 1 TeV, and therefore even
the simplest SUSY model, the Minimal Supersymmetric Standard Model (MSSM, see Ref.~\cite{Martin:1997ns} for a review), predicts a rich phenomenology, including interesting candidates for dark matter (DM) such as the neutralino and sneutrino in 
$R$-parity conserving ($\rp$) models
and the gravitino in
$R$ parity violating  
($\rpv$) models.
However, the LHC started operations several years ago and, with Run 1 already finished, SUSY has not been discovered yet. Because of this, it has been raised the question of whether SUSY is still alive. In our opinion the answer is yes, and we think that there are several arguments in favor of it:

$\bullet$ The lower bounds on SUSY particle (sparticle) masses are smaller or about 1 TeV, depending on the sparticle analyzed. Thus SUSY masses are still reasonable, and in that sense we can remember that the Higgs particle was discovered with a mass close to its (SUSY) perturbative upper bound.

$\bullet$ Because of the complicated parameter space of SUSY, experimentalists use in their analyses simplified models that do not cover the full MSSM. For example, branching-ratio variations are not considered in much detail, and other assumptions are also made.

$\bullet$ Run 2 is going on, and for the moment with a low luminosity of about 20 fb$^{-1}$.
Therefore, to (be prepared) wait for results with higher luminosity seems to be a sensible strategy, since 100 fb$^{-1}$ are expected for the end of Run 2.
 
$\bullet$ Most searches at the LHC assume $\rp$,
with the lightest supersymmetric particle (LSP) stable, requiring therefore missing energy in the final state to claim for detection. However, in the case of
$\rpv$, 
sparticles can decay to standard model particles, and the bounds on their masses become 
weaker.

Nevertheless, despite all these arguments in favour of SUSY, it is honest to recognize that it has its own theoretical problems in the low-energy formulation.
By construction, the MSSM produces too fast proton decay.
In particular, the simultaneous presence of the couplings 
$\lambda'_{ijk} \, L_i \, Q_j \, d_k^c$ and
$\lambda''_{ijk} \, u_i^c\, d_j^c\,  d_k^c$
violating lepton ($L$) and baryon ($B$) number respectively, as well as $\rp$,
would produce this effect.
The usual assumption in the literature of invoking 
$R_p$ 
to avoid the problem, forbidding all $\rpv$ couplings, is perhaps too stringent, since forbidding only one of the above couplings would have been sufficient. We will come back to this point in Section~\ref{rpv}.
So, once eliminated (all) $B$ and $L$ number violating operators, we are left with 
the superpotential of the MSSM:
%
\begin{equation}
W = 
Y^e_{ij} \, \hat H_d\, \hat L_i\, \hat e_j^c 
+
Y^d_{ij} \, \hat H_d\, \hat Q_i \, \hat d_j^c-
Y^u_{ij} \, \hat H_u\, \hat Q_i \, \hat u_j^c +
\mu \,\hat H_u \hat H_d
\ ,
\label{superpotential}
\end{equation}
where $i,j=1,2,3$ are family indexes, 
and
our convention for the contraction of two $SU(2)$ doublets is e.g.
$\hat {H}_u \,  \hat H_d\equiv  \epsilon_{ab} \hat H^a_u \,  \hat H^b_d$,
with $\epsilon_{ab}$ the totally antisymmetric tensor $\epsilon_{12}=1$.

In superpotential~(\ref{superpotential}), the $\mu$ term is necessary e.g. to generate 
Higgsino masses, given the current experimental lower bound of about 100 GeV on chargino masses.
Here we find another problem of SUSY models, the so-called
$\mu$ problem \cite{Kim:1983dt}. 
In the presence of a high-energy theory like a grand unified theory (GUT) or a string theory, with a typical scale of the order of $10^{16}$ GeV or larger, and/or a gravitational theory at the Planck scale, one should be able to explain how to obtain a SUSY mass parameter in the superpotential of the order of the electroweak (EW) scale. 
The MSSM does not solve the $\mu$ problem. One takes for granted that the 
$\mu$ term is there, of the order of the EW scale, and that's it.
In this sense, the MSSM is a kind of effective theory.
Nevertheless,
there are natural 
solutions to this problem that
necessarily introduce new structure beyond the MSSM at low energies. 
Several of these
solutions, and the associated SUSY models, will be discussed below.

The work is organized as follows.
In Section~\ref{mssm}, we will briefly review the popular neutralino DM in the MSSM.
We will also see that the left-handed sneutrino is excluded as candidate for DM from experimental constraints. In Section~\ref{nmssm}, the Next-to-Minimal Supersymmetric Standard Model (NMSSM, see Ref.~\cite{Ellwanger:2009dp} for a review) is introduced as a solution to the $\mu$ problem, and neutralino DM discussed. In an extension of the NMSSM, we will also see that right-handed sneutrino DM is possible.
Finally, in Section~\ref{rpv} we will argue that models with $\rpv$ are viable, solving also the $\mu$ problem.
In particular, the 
`$\mu$ from $\nu$'
Supersymmetric Standard Model
($\mu\nu$SSM~\cite{LopezFogliani:2005yw,Escudero:2008jg}, see Refs.~\cite{Munoz:2009an,Munoz:2016vaa} for reviews),
solves the 
$\mu$-problem through the presence of right-handed neutrino superfields, while
 simultaneously explains the origin of neutrino masses, i.e. in addition it solves the $\nu$ problem.
 Let us emphasize in this sense that in the MSSM, by construction, neutrinos are massless.
Of course, the typical sparticle candidates for DM,
the neutralino or the right-handed sneutrino, 
have very short lifetimes in $\rpv$
models, and can no longer be used as DM.
Nevertheless, the
gravitino 
can be the DM, and
we will discuss its feasibility in the $\mu\nu$SSM, as well as 
its possible detection in gamma-ray satellite experiments such as the
Fermi Large Area Telescope (LAT).



\section{Neutralino DM in the MSSM}
\label{mssm}

As mentioned in the Introduction, the MSSM superpotential in 
Eq.~(\ref{superpotential})
conserves by construction $\rp$. This is a discrete symmetry which assigns quantum number +1 for particles and -1 for sparticles.
As a consequence, 
SUSY particles are produced 
or destroyed only in pairs, and the LSP has to be stable.
This implies that the LSP 
is a possible candidate for DM. In this sense,
it is remarkable that in many regions of the parameter space
of the MSSM
the LSP is the lightest neutralino, 
a physical superposition of the Bino, and neutral Wino and Higgsinos:
\begin{equation}
\tilde\chi^o_1=N_{11}\tilde B^0 + N_{12}\tilde W^0 + N_{13}\tilde H_d^0 + N_{14}\tilde H_u^0
\,.
\label{neutralino}
\end{equation}
Since the neutralino is
an electrically neutral particle, it avoids the problem of charged particles as DM:
they would bind to nuclei and would be excluded from unsuccessful searches for exotic heavy isotopes (see e.g. Ref.~\cite{Kudo:2001ie} and references therein).
Besides, the neutralino is a weakly interacting massive particle (WIMP) and therefore is able to reproduce naturally the amount of relic density that is observed in the Universe, $\Omega_{DM} h^2\sim 1$.
We can then conclude that, in the MSSM, the lightest neutralino 
is a very good
DM candidate (see Ref.~\cite{Munoz:2003gx} for a review).

\begin{figure}[t]
\centering
\includegraphics[scale=0.60]{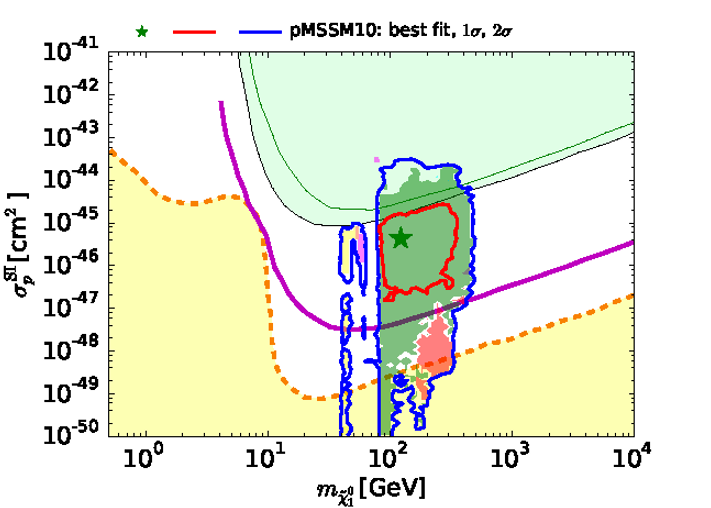}
\caption{The ($m_{\tilde \chi^0_1}$, $\sigma^{\text{SI}}_p$) plane in the pMSSM10.
The green and black lines show the current sensitivities of the XENON100 and LUX experiments, respectively.
The solid purple line show the projected 95$\%$ exclusion sensitivity of the
LUX-Zeplin (LZ) experiment, and the dashed orange line show the astrophysical neutrino `floor', below which astrophysical neutrino backgrounds dominate (yellow region). Figure from \cite{Bagnaschi:2015eha}.}
\label{fig-1}       
\end{figure}

The LHC could produce a neutralino
with a mass of the order of the GeV-TeV.
Such a production and detection would be of course a great success, but not a complete test of the DM theory. 
Even if we are able to measure the mass and interactions of the new particle, checking 
whether the amount of relic density is correct, we would never be able to test if the
candidate is stable on cosmological scales.
A complete confirmation can only arise from experiments where the DM particle is detected as part of the galactic halo or extragalactic structures. 
This can come from
direct and indirect DM searches.
Actually, there has been an impressive progress on this issue in recent years, with significant improvements in the precision and 
sensitivity of experiments.
The combination of LHC data with those provided by direct and indirect searches can be a crucial tool for the identification of the DM.


The neutralino WIMP DM could be detected directly in underground laboratories through its elastic interaction with nuclei inside detectors. 
In view of the LHC1 constraints on SUSY, Higgs data, and flavour physics observables,
in Fig.~\ref{fig-1} 
the current constraints on the parameter space of neutralino DM are shown for the phenomenological MSSM, in which 10 of the effective Lagrangian parameters are treated as independent inputs specified at the EW scale (pMSSM10).
Indirect DM searches of WIMPs are carried out in neutrino and Cherenkov telescopes, and satellites, through the analysis of the DM annihilation or decay products in the Sun, galactic center, galactic halo or extragalactic structures. Such products can be neutrinos, gamma rays and antimatter, and their non-observation put constraints on neutralino DM as well (see Ref.~\cite{Munoz:2012ie} for a review).

\subsection{Sneutrino DM in the MSSM?}

We might wonder whether 
are there other candidates for DM in the MSSM.
In principle, the left-handed sneutrino fulfils the three interesting properties to become a DM candidate~\cite{Ibanez:1983kw,Hagelin:1984wv}. It is a neutral particle, it is stable when it becomes the LSP, and it is a WIMP. However, at the end of the day, it turns out not to be viable for DM. Given its sizable 
coupling to the $Z$ boson, left-handed sneutrinos either annihilate too rapidly,
resulting in a very small relic abundance, or give rise to a large
scattering cross section and are excluded by direct DM
searches.

\section{Is there life beyond MSSM/neutralino DM?}
\label{nmssm}

\subsection{Neutralino DM in the NMSSM}

The NMSSM
provides an elegant solution to the $\mu$ problem of the MSSM via the
introduction of a singlet superfield $\hat S$ under the standard model gauge group. 
Substituting now the $\mu$-term in (\ref{superpotential}) by
\begin{equation}
\begin{array}{rcl}
W  = 
\lambda\ \hat S \hat H_u \hat H_d\ + k\ \hat S\hat
S\hat S\,,
\end{array}\label{superpotentialnmssm}
\end{equation}
%
when the scalar component of the superfield $\hat S$,
denoted by $S$, acquires
a vacuum expectation value (VEV) of order the SUSY breaking scale, 
an effective interaction $\mu \hat H_1 \hat H_2$ is generated
through the first term in  (\ref{superpotentialnmssm}), with  
$\mu\equiv\lambda \langle S\rangle$.
This
effective coupling is naturally of order the EW scale if the SUSY  breaking
scale is not too large compared with $M_W$, as expected. In fact, in the NMSSM 
the EW scale
exclusively originates from the SUSY-breaking scale. 
The second term in (\ref{superpotentialnmssm}) 
is allowed by the gauge symmetry,
and avoids, as the $\mu$-term in the MSSM, 
the existence of a Goldstone boson.

Due to the presence of the superfield $\hat S$,
in addition to the MSSM fields, the NMSSM contains an extra scalar and pseudoscalar in the Higgs sector, as well as an additional singlino/neutralino. These new fields
mix with the corresponding MSSM ones, giving rise to a richer and more complex
phenomenology.
For example, the results concerning the 
possible detection of neutralino DM turn out to be modified 
with respect to those of the MSSM in regions of the parameter space.

\subsection{Sneutrino DM in an extended NMSSM 
}
\label{extended}

An interesting extension of the NMSSM can help to explain
the origin of neutrino masses.
Since experiments induce us to introduce
right-handed neutrino superfields, 
superpotential (\ref{superpotentialnmssm})
can be extended with \cite{Kitano:1999qb}:
\begin{equation}
\begin{array}{rcl}
\delta W  = 
Y^{\nu}_{ij}\ \hat H_u\, \hat L_i \, \hat \nu^c_j 
+ \kappa_{ij}\ \hat S \hat \nu^c_i\hat \nu^c_j\,.
\end{array}\label{deltasuperpotentialMajorana2}
\end{equation}
%
Majorana masses for right-handed neutrinos of the order of the EW scale are generated dynamically
through the VEV of the singlet $S$, $M_{\nu}=\kappa \langle S\rangle$. This is an example of 
a seesaw at the EW scale.
Light masses are then obtained with a value 
$m_{\nu} \simeq Y_{\nu}^2 v_u^2/M_{\nu}$, 
which implies Yukawa couplings $Y_{\nu}\lsim$ $10^{-6}$, i.e. of the same order as the electron Yukawa.


As discussed above in the context of the MSSM, the left-handed sneutrino cannot be used as a DM candidate. Actually, a purely right-handed sneutrino either in a natural way, because of its very weak couplings with the rest of the matter implying a scattering cross section too small (supressed by $Y^{\nu}$), and a relic density too large.
However, now,
through its direct coupling to the singlet in (\ref{deltasuperpotentialMajorana2}), the right-handed sneutrino
can be not only a thermal relic DM, but
  also have a large enough scattering cross section with nuclei as to be
  detected (see~\cite{Cerdeno:2008ep} 
and~\cite{Cerdeno:2014cda,Cerdeno:2015ega}, and references therein).



\section{Is there life beyond $\rp$/neutralino-sneutrino DM} 
\label{rpv}

As discussed in the Introduction, to impose $\rp$ in SUSY models 
may be too stringent, since the $\rpv$ couplings which are harmless for proton decay would also be forbidden.
A less drastic solution, taking into account that the choice of 
$R_p$ 
is {\it ad hoc},
is to use other $Z_N$ discrete symmetries to forbid only
$\lambda''_{ijk}$. This is the case e.g. of $Z_3$ Baryon-parity~\cite{Ibanez:1991pr} which also prohibits dimension-5 proton decay operators, unlike $R_p$.
In addition, this strategy seems reasonable if one expects all discrete symmetries to arise from the breaking of gauge symmetries of the underlying unified theory,
because
Baryon-parity and 
$R_p$ are the only two generalized parities which are `discrete gauge' anomaly free.
Actually, this can occur in string compactifications where
the matter superfields can have
several extra $U(1)$ charges broken spontaneously at high energy, and as a consequence residual $Z_N$ symmetries 
are left in the low-energy theory.
The same result can be obtained by the complementary mechanism pointed out in 
Ref.~\cite{Escudero:2008jg}, that stringy selection rules can naturally forbid the $\lambda''_{ijk}$ couplings since matter superfields are located in general in different sectors of the compact space.

The gravitino turns out to be an interesting candidate for DM in $\rpv$ 
models. It has
an interaction term in the supergravity Lagrangian
with 
the photon and the photino. Since
the photino and the neutrinos are mixed in the neutral fermion mass matrix due to the $\rpv$,
the gravitino will be able to decay 
into a photon and a neutrino. Nevertheless, this decay is suppressed both by
the gravitational interaction (the gravitino is a superWIMP) and by the small $\rpv$ coupling, making
the gravitino lifetime much longer than the age of the Universe~\cite{Takayama:2000uz}.
Adjusting the reheating temperature one can also
reproduce the correct relic density.

\subsection{Gravitino DM in the $\mu\nu$SSM}
\label{munussm}

Right-handed neutrinos are likely to exist in order to generate neutrino masses.
Then, given the
fact
that sneutrinos are allowed to get VEVs,
we may wonder why not to use $\rpv$ terms
of the type $\hat \nu^c \hat H_u\hat H_d$ 
to produce an effective  $\mu$ term.
This would allow us to solve the $\mu$ problem of the MSSM, 
without having to introduce an extra singlet superfield
as in case of the NMSSM.
This is the basic idea of the $\mu\nu$SSM~\cite{LopezFogliani:2005yw,Escudero:2008jg}: natural particle content 
without the $\mu$ problem.
Thus, in addition to the MSSM Yukawa couplings for quarks and charged leptons, 
the
$\mu$$\nu$SSM superpotential contains: 
%
\begin{equation}
W  = 
-Y^\nu_{ij} \, \hat H_u\, \hat L_i \, \hat \nu^c_j 
 + \lambda_{i} \,\hat H_u \hat H_d \, \hat \nu^c_i
+\frac{1}{3}
\kappa_{ijk} 
\hat \nu^c_i\hat \nu^c_j\hat \nu^c_k\ .
\label{superpotentialmunussm}
\end{equation}
%
When the scalar components of the superfields $\hat\nu^c_i$,
denoted by $\tilde\nu_{iR}^*$, acquire
VEVs of order the EW scale, 
an effective interaction $\mu \hat H_u \hat H_d$ is generated
through the second term in (\ref{superpotentialmunussm}), with  
$\mu\equiv
\lambda_i \langle \tilde \nu_{iR} \rangle^*$.
The third term in (\ref{superpotentialmunussm}) 
is allowed by all symmetries,
and avoids the presence of a Goldstone boson associated to a global $U(1)$
symmetry, similarly to the case of the NMSSM.
In addition, it contributes to generate 
effective Majorana masses for neutrinos at
the EW scale
$M_{ij}=\sqrt 2\kappa_{ijk} \langle \tilde \nu_{kR} \rangle^*$, which together with the
Dirac masses generated by the first term, produce correct neutrino masses.
Thus, the $\mu$$\nu$SSM solves the $\mu$ and the $\nu$ problems,
by simply introducing right-handed neutrinos.

Since the gravitino decays producing a monochromatic photon with an energy half of the gravitino mass,
the prospects for detecting these $\gamma$ rays in satellite experiments can be very interesting, and therefore it seems important to know $\mn$ predictions concerning gravitino DM detection, first studied in Ref~\cite{Choi:2009ng}.
In recent works~\cite{Albert:2014hwa,Gomez-Vargas:2016ocf}, a complete analysis of the detection of $\mn$ gravitino dark matter through 
$\gamma$-ray observations was carried out. 
In addition to the two-body decay producing an
anisotropic sharp line,
the three-body decays producing a smooth spectral signature were included in the analysis. First, 
a deep exploration of the low-energy parameter space of the $\mn$ was performed, taking into account that neutrino data must be reproduced. 
Then, the $\gamma$-ray fluxes predicted by the model were compared with \Fermi LAT observations.
In particular, with the
95$\%$ CL upper limits on the total diffuse extragalactic $\gamma$-ray background using 50 months of data, together with the upper limits on line emission from an updated analysis using
69.9 months of data.
For standard values of bino and wino masses,
gravitinos with masses larger than 4 GeV, or lifetimes smaller
than $10^{28}$ s, produce too large fluxes and are excluded as DM candidates. 
However, when limiting scenarios with large and close values of the gaugino masses are considered,
the constraints turn out to be less stringent, excluding masses larger than 17 GeV and lifetimes smaller
than $4\times 10^{25}$ s.


\vspace{0.5cm}

\noindent {\bf Acknowledgments} 

\noindent 
This work was supported
by grants FPA2015-65929-P MINECO/FEDER UE, Consolider-Ingenio 2010 MultiDark CSD2009-00064, and SEV-2012-0249 `Centro de Excelencia Severo Ochoa'.


%
 \bibliography{biblio}
%

\end{document}